\documentclass[twoside,11pt]{article}

\usepackage{blindtext}

\usepackage{algorithm}
\usepackage{algpseudocode}
\usepackage{amsmath}
\usepackage{tikz}
\usepackage{wrapfig}

\usepackage{jmlr2e}

\usepackage{lastpage}

\ShortHeadings{PDFA Distillation via String Probability Queries}{PDFA Distillation via String Probability Queries}
\firstpageno{1}

\begin{document}

\title{PDFA Distillation via String Probability Queries}

\author{\name Robert Baumgartner \email r.baumgartner-1@tudelft.nl \\
       \name Sicco Verwer \email s.e.verwer@tudelft.nl \\
       \addr Department for Electrical Engineering, Mathematics, and Computer Science \\
       Algorithmics Group \\
       Technical University of Delft, Netherlands}


\maketitle

\begin{abstract}
Probabilistic deterministic finite automata (PDFA) are discrete event systems modeling conditional probabilities over languages: Given an already seen sequence of tokens they return the probability of tokens of interest to appear next. These types of models have gained interest in the domain of explainable machine learning, where they are used as surrogate models for neural networks trained as language models. In this work we present an algorithm to distill PDFA from neural networks. Our algorithm is a derivative of the L\textsuperscript{\#} algorithm and capable of learning PDFA from a new type of query, in which the algorithm infers conditional probabilities from the probability of the queried string to occur. We show its effectiveness on a recent public dataset by distilling PDFA from a set of trained neural networks. 
\end{abstract}


\section{Introduction}

Neural networks (NNs) are a powerful means for sequence modeling, in which they have gained strong interest from both commercial companies as well as the research community. A commonly recognised drawback to their expressive power is their inherent difficulty for humans to understand their decision processes (see e.g. \cite{black_box_survey}). Attempts to explain neural networks that are trained for sequence modeling, returning the conditional probability $P(a|x)$ over a corpus $a \in \Sigma$ and $x \in \Sigma^*$, have led to several works in distilling probabilistic finite automata (PFA) and probabilistic deterministic finite automata (PDFA) as surrogate models from such NNs \citep{weiss_2, okudono, spectral_distillation, mayr_1, mayr_2}. Advantages of PFA and PDFA are that, once distilled, they provide a computationally cheap model compared with the actual neural networks. Moreover, they are naturally visualizable and can be interpreted by an individual with expert knowledge, and can in theory be learned under PAC constraints \citep{pac_bounds_clark_thollard}.

Algorithms to distill PFAs and PDFAs from sequence modeling NNs all work by asking the NN conditional probabilities $P(a|x)$. There are 3 kinds of approaches taken so far: The first group \citep{weiss_2} adapts the L\textsuperscript{*} algorithm by constructing a real table made of the responses of the network. The main difference to the L\textsuperscript{*} algorithm's observation table \citep{l_star} is that this table contains real valued numbers rather. Then a notion of similarity over rows is defined, and the automaton is constructed similar to L\textsuperscript{*}. The second group constructs a similar kind of table, but uses spectral learning \citep{spectral_learning_balle} to extract an automaton from this table \citep{okudono, spectral_distillation}. A drawback of this group is that the resulting models are non-deterministic, making them hard to be interpreted. Recently a new group emerged, building an observation tree from the queries rather than a table, and subsequently minimizing this tree \citep{mayr_1, mayr_2}.

We want to build on the idea to learn on the observation tree and add another type of query that can be asked. Instead of asking the next symbol probabilities $P(a|x)$, we ask the full string probability $P(x a)$ directly. We employ an observation tree representing the answers and probabilities to infer the conditional probabilities $P(a|x)$ and use the inferred probabilities to minimize the tree. We employ a merge heuristic that enforces an error bound $\mu>0$ during our minimization procedure to make sure predicted probabilities for each already seen sequence $x$ stay within those error bounds. We evaluate our algorithm on the TAYSIR competition dataset \citep{taysir}. All code is made available through our own public repository\footnote{\url{https://github.com/tudelft-cda-lab/FlexFringe}}. A mathematical motivation for how we infer probabilities is provided in the appendix \ref{sec:proofs}.

\section{PDFAs}

A PDFA is typically depicted via a 5-tuple $\mathcal{A} = \{q_0, Q, \Sigma, \tau, \pi \}$, where $q_0$ represents the unique starting state, and $Q$ denotes the set of all states $q$. $\Sigma$ is a finite set of tokens, and an individual token is written shorthand via $a$. $x$ denotes an arbitrary string over $\Sigma^*$. $\Sigma^*$ is the set of all possible strings over $\Sigma$ with finite length, and we write $|x|$ to denote the length of string $x$. Concatenations of strings are simply written in the form of $a_0a_1...$ for strings of tokens, or $ax$ and $xa$ for a symbol $a$ preceding and following a string $x$ respectively. $\lambda$ denotes the empty string with length $|\lambda|=0$, and $\lambda x = x = x \lambda$. 

Traversing the automaton is done via $\tau : Q \times \Sigma \rightarrow Q$. $\tau$ can be recursively defined via $\tau(q, \lambda)=q$ and $\tau(q, ax)=\tau(\tau(q, a), x)$. We say that a state $q'$ is reachable from state $q$ if there exists at least one string $x$ s.t. $q'=\tau(q, x)$. We write in short $Q_{\tau(q)}$ to denote the set of all states reachable from $Q$, and we call $X_{\tau(q)}$ the set of shortest strings to reach them. A shortest string $x$ in $X_{\tau(q)}$ is a string s.t. $q'=\tau(q, x)$ and there is no string $x'$ s.t. $q'=\tau(q, x')$ and $|x'|<|x|$. We call a PDFA an observation tree $\mathcal{OT}$ iff each state $q$ has a unique and only one sequence $x_q^A$ s.t. $q = \tau(x_q^A)$. We call $x_q^A$ the access sequence of $q$. We note that on an observation tree $\mathcal{OT}$ $X_{\tau(q)}$ is clearly defined for each state of the tree.

\begin{wrapfigure}{R}{0.5\textwidth}\label{fig:pdfa}
\begin{tikzpicture}[scale=0.2]
\tikzstyle{every node}+=[inner sep=0pt]
\draw [black] (31.7,-19.7) circle (3);
\draw (31.7,-19.7) node {$q_1/0.3$};
\draw [black] (46.4,-19.7) circle (3);
\draw (46.4,-19.7) node {$q_2/0.1$};
\draw [black] (16,-19.7) circle (3);
\draw (16,-19.7) node {$q_0/0.1$};
\draw [black] (9.4,-19.7) -- (13,-19.7);
\fill [black] (13,-19.7) -- (12.2,-19.2) -- (12.2,-20.2);
\draw [black] (14.677,-17.02) arc (234:-54:2.25);
\draw (16,-12.45) node [above] {$a/0.3$};
\fill [black] (17.32,-17.02) -- (18.2,-16.67) -- (17.39,-16.08);
\draw [black] (18.646,-18.298) arc (111.79563:68.20437:14.017);
\fill [black] (29.05,-18.3) -- (28.5,-17.54) -- (28.13,-18.46);
\draw (23.85,-16.8) node [above] {$b/0.6$};
\draw [black] (29.243,-21.407) arc (-62.55387:-117.44613:11.701);
\fill [black] (18.46,-21.41) -- (18.94,-22.22) -- (19.4,-21.33);
\draw (23.85,-23.22) node [below] {$a/0.2$};
\draw [black] (45.077,-17.02) arc (234:-54:2.25);
\draw (46.4,-12.45) node [above] {$a/0.2,\mbox{ }b/0.7$};
\fill [black] (47.72,-17.02) -- (48.6,-16.67) -- (47.79,-16.08);
\draw [black] (34.7,-19.7) -- (43.4,-19.7);
\fill [black] (43.4,-19.7) -- (42.6,-19.2) -- (42.6,-20.2);
\draw (39.05,-20.2) node [below] {$b/0.5$};
\end{tikzpicture}

\caption{An example of a PDFA consisting of three states. $q_0$ is the initial states, and the stopping probabilities are $\pi(q_0)=0.1$, $\pi(q_1)=0.3$, and $\pi(q_2)=0.1$.}
\end{wrapfigure}
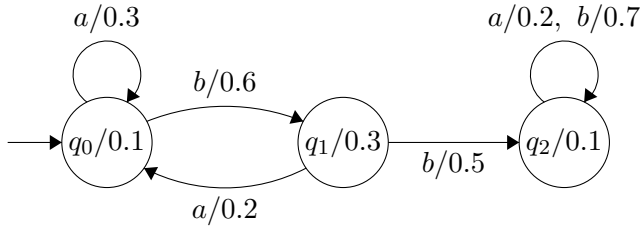

Finally $\pi: Q\times \Sigma \rightarrow [0, 1]$ and $\pi: Q \rightarrow [0, 1]$ assigns probabilities to transitions within the automaton. We call $\pi(q)$ the stopping probability of state $q$, denoting the event that a string reaches $q$ with its last possible transition. A PDFA requires $\forall q \in Q: \text{ } \sum_{a\in\Sigma} \pi(q, a) + \pi(q) = 1$. The probability of a string $x=a_0a_1a_2...a_n$ can then be computed via $\pi(x)=\pi(q_0, a_0)\cdot \pi(\tau(q_0, a_0), a_1)\cdot...\cdot \pi(\tau(q_{n-1}, a_{n-1}), a_n) \cdot \pi(q_{n+1})$. Note here that we introduced shorthand notation for the string probability as $\pi(x)$. Fig. \ref{fig:pdfa} depicts an easy example of a PDFA.

\newpage

\section{Learning algorithm}\label{sec:learning_algorithm}

In adaptation of the MAT framework \citep{l_star} our algorithm distinguishes two abstract entities: The first one is the system under learning (SUL), abstracted away by the teacher, and the learner, whose goal it is to find a model representative of the SUL. We assume the teacher to provide answers of the form $P(x)$: Given a string $x$, it returns the probability for $x$ to occur. Furthermore, the learner can ask the teacher for equivalence: Given a hypothesis $\mathcal{H}$, does it model the behavior of the SUL $\mathcal{T}$ sufficiently close? 

The core idea of our learner is to build an observation tree that stores the observed probabilities. The algorithm propagates new string probabilities throughout relevant branches of the observations tree to compute the already seen probability mass in each node, and periodically updates $\pi$ with the new estimates. We first explain the observation tree and how the algorithm constructs it, and then explain how it turns it into a hypothesis.

\subsection{Growing the observation tree}

In order for the learner to find a surrogate model for the SUL it generates one or more hypotheses and tests each of them. Similar to the L\textsuperscript{\#} algorithm \citep{l_sharp} it first builds an observation tree $\mathcal{OT}$, representing the set of already asked input strings, as well as the answers the teacher provided. Initially the observation tree consists of a single node, which will be $q_0$, $x_{q_0}^A=\lambda$. The tree is then grown from $q_0$ by asking queries and creating respective new nodes, which we encapsulate in the $ExtendFringe$-operation (Alg. \ref{alg:extend_fringe}). Upon creating a node within the observation tree, the learner assigns it four attributes:

\begin{enumerate}
    \item A field for modeling the probabilities $\pi(q, a)$. 
    \item A field for modeling the probability $\pi(q)$.
    \item An attribute for saving the probability $P(x_1^A)$, which we call the access probability of $q$ and write $P_A(q)$.
    \item A weight attribute used to estimate probabilities for each $a\in\Sigma$, denoted by $m(q, a)$.
\end{enumerate}

Initialization of a node is done according to Alg. \ref{alg:init_node}, which initializes its attributes. Additionally, each time a new node is created and initialized, it provides better estimates for the probabilities of nodes that are part of $T(x_q^A)$. Therefore, each time a new node is created and updated, these nodes are updated through the subroutine depicted in Alg. \ref{alg:update_path}. At certain points of the algorithm it needs to update $\pi(q)$ and $\pi(q, a)$. It does this via Alg. \ref{alg:normalize_node}. Our goal is to keep the probabilities $\pi(x_q^A)$ in the observation tree precise. A depth-first-search (DFS) (Alg. \ref{alg:dfs_update}) routine ensures this condition. We provide mathematical motivation for our operations in appendix \ref{sec:proofs}. 

\begingroup
\renewcommand{\algorithmicrequire}{\textbf{Input:}}
\renewcommand{\algorithmicensure}{\textbf{Output:}}

\begin{algorithm}[t]
\caption{Extend fringe}\label{alg:extend_fringe}
\begin{algorithmic}
\Require Set of fringe nodes $\mathcal{F}$, alphabet $\Sigma$
\Ensure Set of new fringe nodes $\mathcal{F}_n$

\State $\mathcal{F}_n \leftarrow$ empty set

\ForAll{Nodes $q$ in $\mathcal{F}$}
    \ForAll{$a \in \Sigma$}
        \State Create node $q'$ satisfying $q'=\tau(q, a)$
        \State $InitializeNode(q')$
        \Comment{Algorithm \ref{alg:init_node}}
        \State $\mathcal{F}_n \leftarrow \mathcal{F}_n \cup \{q'\}$
    \EndFor
\EndFor

\Return $\mathcal{F}_n$

\end{algorithmic}
\end{algorithm}

\endgroup

\begingroup
\renewcommand{\algorithmicrequire}{\textbf{Input:}}
\renewcommand{\algorithmicensure}{\textbf{Output:}}

\begin{algorithm}
\caption{Initialize node}\label{alg:init_node}
\begin{algorithmic}
\Require Node $q$, alphabet $\Sigma$

\State $x_q^A \leftarrow q.getAccessString()$
\Comment{Can be stored or computed}
\State $\pi(q) \leftarrow P(x^A_q)$
\State $P_A(q) \leftarrow P(x^A_q)$

\ForAll{$a \in \Sigma$}
    \State $m(q, a) \leftarrow P(x^A_q a)$
    \State $\pi(q, a) \leftarrow P(x^A_q a)$
\EndFor

\State $UpdatePath(q, x_q^A, P(x_q^A))$
\Comment{Algorithm \ref{alg:update_path}}

\end{algorithmic}
\end{algorithm}

\begin{algorithm}
\caption{Update path}\label{alg:update_path}
\begin{algorithmic}
\Require Node $q$, string $x_q^A$, probability $P(x_q^A)$

\State $\sigma \leftarrow \lambda$
\ForAll{$i \leftarrow 0$ up to $|x_q^A|$}
    \State $a \leftarrow a_i$
    \Comment{$x_q^A=a_0 a_1 a_2 ...$}
    \State $q' \leftarrow \tau(\sigma)$
    \State $m(q', a) \leftarrow m(q', a) + P(x_q^A)$
    \State $\sigma \leftarrow \sigma a$
\EndFor

\end{algorithmic}
\end{algorithm}

\begin{algorithm}[t]
\caption{DFS update}\label{alg:dfs_update}
\begin{algorithmic}
\Require Root node $q_0$, alphabet $\Sigma$

\State $S_Q \leftarrow$ empty stack
\State $S_Q.push(q_0)$
\State $S_P \leftarrow$ empty stack
\State $S_P.push(1)$

\While{$S_Q$ not empty}
    \State $q \leftarrow S_Q.pop()$
    \State $p \leftarrow S_P.pop()$
    \State $\pi(q) \leftarrow \frac{P_A(q)}{p}$
    \Comment{Guarantees that $\pi(x_q^A)$ be correct for all $q$ in tree}
    \State $NormalizeNode(q)$
    \Comment{Algorithm \ref{alg:normalize_node}}
    \ForAll{$a\in \Sigma$}
        \If{$\tau(q, a)$ defined}
            \State $S_Q.push(\tau(q, a))$
            \State $S_P.push(p \cdot \pi(q, a))$
        \EndIf
    \EndFor
\EndWhile

\end{algorithmic}
\end{algorithm}

\begin{algorithm}
\caption{Normalize node}\label{alg:normalize_node}
\begin{algorithmic}
\Require Node $q$, alphabet $\Sigma$

\State $s \leftarrow 0$
\ForAll{$a\in \Sigma$}
    \State $s \leftarrow s + m(q, a)$ 
\EndFor

\State $f \leftarrow \frac{1-\pi(q)}{s}$
\ForAll{$a\in \Sigma$}
    \State $\pi(q, a) \leftarrow f \cdot m(q, a)$ 
\EndFor

\end{algorithmic}
\end{algorithm}

\endgroup

\subsection{Finding a hypothesis candidate}\label{sec:finding_hypothesis}

Just like the L\textsuperscript{\#} algorithm we employ the red-blue-framework \citep{lang_1998} when growing the observation tree. We separate the tree into three distinct parts: A core of identified red nodes, a fringe of blue nodes, and a set of white nodes that are neither red nor blue. Initially, the algorithm starts with only a red state $q_0$, which it adds a fringe to through Alg. \ref{alg:extend_fringe}. These newly created nodes are marked blue. A blue node is a node which is not red, but whose parent node is red. Then the algorithm triggers the DFS routine to estimate the probabilities and tries to minimize the automaton. 

We utilize techniques from state machine learning by proposing a merge check. We say that a pair of a red node $q_r$ and a blue node $q_b$ are consistent under a threshold $\mu \in [0, 1)$ iff 

\begin{equation}\label{eq:error_bound_check}
    d(q_r, q_b) = \left| \frac{\pi(x^A_{q_b})}{\pi(q_b)} \cdot \pi(q_r) - P(x^A_{q_b}) \right| \leq \mu.
\end{equation}

In order to build deterministic models we perform a determinization process whenever we merged two states (explained in e.g. \citeauthor{real_time_automata}). We call $q_r$ and $q_b$ mergeable iff $\forall q_b' \in Q_{\tau(q_b)}$ it holds that either $\tau(q_r, x)$, $x \in X_{q_b}$ is not defined, or $q_r'=\tau(q_r, x)$ and $d(q_r', q_b')\leq \mu$. This way we ensure that our error bound holds for all strings the algorithm has seen so far. We note that the red-blue-framework ensures that the nodes $Q_{\tau(q_b)}$ form a tree, thus do not form loops. The resulting access sequence of a merge is $x_{q_r}^A$.

Every time the algorithm minimizes the observation tree the algorithm assumes a red root node $q_0$, a fringe of blue nodes $q'$ s.t. $q'=\tau(q_0, a) \text{ } \forall a \in \Sigma$, and a possibly empty set of white nodes\footnote{We guarantee these starting conditions through the algorithm, a description follows later.}. Goal of the algorithm is to find a \textit{complete basis} $\mathcal{B}$, i.e. a set of red nodes $\mathcal{B}$ s.t. $\forall q \in \mathcal{B} \text{, } \forall a \in \Sigma: \text{ } \tau(q, a) \in \mathcal{B}$. This is done by minimizing the observation tree through an iterative procedure, where each iteration processes the current set of blue nodes. To do this the algorithm compares each of the currently blue nodes with each of the currently red nodes and chooses an operation for each blue node. The algorithm remembers the chosen operation for each node and performs these at the end of the operation. If a blue node can be merged with a red node it will remember this merge as the operation. If a blue node can be merged with multiple red nodes the merge that introduces minimal error according to Eq. \ref{eq:error_bound_check} will be preferred. We consider a blue node that cannot be merged with any current red node as a new identified state and turn it red. We describe the minimization of one layer in Alg. \ref{alg:merge_layer}.

\begin{algorithm}
\caption{Merge layer}\label{alg:merge_layer}
\begin{algorithmic}
\Require Set of red nodes $\mathcal{R}$, set of blue nodes $\mathcal{B}$, threshold $\mu$

\State $H_O \leftarrow$ empty hash-table
\Comment{Stores the optimal operations for each blue node}

\ForAll{Nodes $q \in \mathcal{B}$}
    \State $s_{min} \leftarrow 0$
    \Comment{If multiple merges possible use $s_{min}$ to select one}
    \ForAll{Nodes $q' \in \mathcal{R}$}
        \If{$q$ and $q'$ mergeable under $\mu$ and $score(q, q')< s_{min}$}
        \Comment{$score(q, q')$: Eq. \ref{eq:error_bound_check}}
            \State $H_O.insert(q, merge(q, q'))$ 
            \State $s_{min} \leftarrow score(q, q')< s_{min}$
        \EndIf
    \EndFor
    \If{No operation in $H_O$ for $q$}
    \Comment{True if $q$ could not be merged}
        \State $H_O.insert(q, turn\_red(q))$
    \EndIf
\EndFor

\ForAll{Nodes $q\in \mathcal{B}$}
    \State $o \leftarrow H_O.search(q)$
    \State Perform $o$
\EndFor

\end{algorithmic}
\end{algorithm}

If every blue node of one layer was mergeable with at least one red node the resulting set of red nodes will form a complete basis. In this case the current model is forwarded to the hypothesis testing procedure. If all layers of the current automaton have been merged and no complete basis has been found the algorithm resets all performed operations to retrieve the original observation tree again and extends the fringe. This procedure is repeated until a complete basis has been found or until an early stop criterion has been reached, see appendix \ref{sec:practical_consideration}.

\subsection{Hypothesis testing and counterexample processing}\label{sec:testing_and_counterexample_processing}

Once a valid hypothesis $\mathcal{H}$ has been found the learner asks the teacher for equivalence (see section \ref{sec:learning_algorithm}). We consider $\mathcal{T}$ and $\mathcal{H}$ equivalent iff $\forall x \in \Sigma^*: |P(x)-\pi(x)|\leq \mu$. 

Asking the equivalence query can have two possible outcomes: 1. The teacher deems $\mathcal{H}$ and $\mathcal{T}$ to be equivalent. In this case it responds with `yes', and the learner returns $\mathcal{H}$ and terminates the algorithm. 2. The teacher deems $\mathcal{H}$ and $\mathcal{T}$ not equivalent. In this case it responds with `no' and returns a counterexample $x_{cex}$ to the learner s.t. $|P(x_{cex}) - \pi(x_{cex})| > \mu$. In the latter case the learner gets the chance to process the counterexample to improve $\mathcal{H}$. 

We follow a simple strategy. We reset $\mathcal{H}$ to the original observation tree, and subsequently parse the tree via $\tau(x_{cex})$. Because we are guaranteed that all previously asked strings do not violate our equivalence condition (see subsection \ref{sec:finding_hypothesis}), parsing via $\tau(x_{cex})$ will eventually reach a node $q_i$ s.t. $\tau(q_i, x_{cex, i})$ will not be defined. Here, $x_{cex, i}$ is the $i$-th token of $x_{cex}$. In order to process the counterexample the learner iteratively creates nodes s.t. $\tau(q_j, x_{cex, j})\text{, } j \geq i$ is defined, and initializes them via Alg. \ref{alg:init_node}. Then, the learner continues searching for a complete basis. The overall algorithm flow is depicted in Alg. \ref{alg:main}.

\begin{algorithm}
\caption{Main routine}\label{alg:main}
\begin{algorithmic}
\Require Threshold $\mu$, alphabet $\Sigma$, access to SUL $\mathcal{T}$ via teacher

\State $\mathcal{H} \leftarrow q_0$
\State $\mathcal{F} \leftarrow \{ q_0 \}$
\Comment{$\mathcal{F}$ is set of fringe nodes}
\While{No termination signal}
\Comment{Explained in appendix \ref{sec:early_stop}}
    \State $\mathcal{F} \leftarrow ExtendFringe(\mathcal{F}, \Sigma)$
    \While{Blue nodes $\mathcal{B}$ remain and no complete basis found}
        \State $\mathcal{R} \leftarrow$ set of currently red nodes
        \State $MergeLayer(\mathcal{R}, \mathcal{B}, \mu)$
    \EndWhile

    \If{Complete basis found}
        \State Ask teacher equivalence query
        \If{$\mathcal{H}$ and $\mathcal{T}$ equivalent}
            \Return $\mathcal{H}$
        \EndIf
        \State Reset $\mathcal{H}$ to observation tree
        \State Process counterexample $x_{cex}$
        \Comment{See section \ref{sec:testing_and_counterexample_processing}}
    \Else 
        \State Reset $\mathcal{H}$ to observation tree
    \EndIf
\EndWhile

\end{algorithmic}
\end{algorithm}

\section{Experiments and results}

We tested our algorithm on the TAYSIR competition \cite{taysir}. For obvious reasons we focused on the part of the dataset that allows the inference of PDFA, namely track 2. We implemented our algorithm in Flexfringe\footnote{\url{https://github.com/tudelft-cda-lab/FlexFringe}}, and ran it on the respective models. In order to keep the models small and concise, but also for faster inference, we kept the hyperparameter $\mu$ relatively large at a value of $0.0001$. All experiments have been run on a notebook with Ubuntu 22.04 64-bit, 16GB RAM, and an Intel i7 CPU @ 2.60GHz x 12. The maximum depth that we explored was set to $6$.

Table \ref{tab:taysir_track_2_results} shows the results the we achieved. Here we show the scenario of track 2 of the competition, mention the size of the alphabet, and compare the size of the resulting PDFA and the achieved mean-squared-error (MSE) with the winners of the competition, who also used automata learning. Additionally, we provide the run-times in table \ref{tab:taysir_track_2_runtimes}. 

We can see that we achieve very low MSE already with very few states, indicating that these PDFA are already capable of modeling the language well. The time an inferred observation tree reached the depth of $6$ was scenario 10, which explains its larger run-time.

\begin{table}[!t]
\centering
\begin{tabular}{ ||c|c||c|c||c|c|| }
 \hline
 Scenario & $|\Sigma|$ & MSE Winners & MSE pL\textsuperscript{\#} & $n$ Winners & $n$ pL\textsuperscript{\#} \\
 \hline\hline
 1 & 33 & $0.175e^{-6}$ & $0.679e^{-6}$ & 866 & 21 \\ 
 \hline
 2 & 20 & $0.97e^{-8}$ & $0.645e^{-5}$ & 131 & 91 \\ 
 \hline
 3 & 7 & $0.3e^{-10}$ & $0.749e^{-6}$ & 110 & 30 \\
 \hline
 4 & 15 & $0.6e^{-11}$ & $0.146e^{-7}$ & 105 & 10 \\ 
 \hline
 5 & 20 & $0.7^{-13}$ & $0.684e^{-7}$ & 123 & 11 \\ 
 \hline
 6 & 33 & $0.1971e^{-6}$ & $0.843e^{-6}$ & 318 & 20 \\ 
 \hline
 7 & 20 & $0$ & $0.364e^{-10}$ & 170 & 16\\ 
 \hline
 8 & 66 & $0.443e^{-7}$ & $0.123e^{-5}$ & 162 & 16 \\ 
 \hline
 9 & 20 & $0$ & $0.674e^{-10}$ & 55 & 17 \\ 
 \hline
 10 & 33 & $0.124e^{-6}$ & $0.21e^{-4}$ & 1412 & 38 \\ 
 \hline
\end{tabular}
\caption{Results on track 2 of the TAYSIR competition set. In reference to the L\textsuperscript{\#} algorithm we call our method pL\textsuperscript{\#} here. $n$ indicates the number of states of the resulting hypothesis model.}
\label{tab:taysir_track_2_results}
\end {table}

\begin{table}[!t]
\centering
\begin{tabular}{ ||c|c|c|c|c|c|c|c|c|c|| }
 \hline
 1 & 2 & 3 & 4 & 5 & 6 & 7 & 8 & 9 & 10 \\
 \hline
 6m31s & 10m9s & 11m8s & 2m57s & 29m22s & 6m19s & 5m24s & 8m30s & 5m32s & 105m56s \\
 \hline
\end{tabular}
\caption{Run-times of the experiments conducted.}
\label{tab:taysir_track_2_runtimes}
\end {table}

\section{Discussion and conclusion}

In this work we presented a novel active automata learning algorithm. Inspired by the L\textsuperscript{\#} algorithm it learns directly on an observation tree. Capable of asking a new type of query, namely string probability queries, it learns PDFA as a surrogate model of a system under learning. We showed the effectiveness of the algorithm experimentally by distilling PDFA from trained neural networks. Using our method we achieved low inference errors with relatively few states already. We provide mathematical motivation as well as convergence behavior of the inferred probabilities. A possible application of our algorithm lies in reverse engineering, where we can infer next-symbol-probabilities from only whole-string-probabilities when next-symbol-probabilities are not available.

\newpage

\acks{This work is supported by NWO TTW VIDI project 17541 - Learning state machines from infrequent software traces (LIMIT).}

\appendix

\section{Practical considerations}\label{sec:practical_consideration}

\subsection{Early stopping criteria}\label{sec:early_stop}

Under normal circumstances the algorithm will terminate and return a hypothesis once it found a complete basis for which the teacher cannot determine a counterexample. In case the SUL is very complicated however the algorithm can potentially run for a very long time. In this case early stop criteria are desired. We introduced a maximum number of $ExtendFringe$-operations the algorithm is allowed to do, which is similar to a maximum depth that the observation tree can grow. An exception is given by the counterexample processing routine, which can grow paths deeper than this depth. Other methods would be to limit the time the learner runs, or to set a limit on the probability mass covered by the membership queries. Whenever such a criterion is met the algorithm returns an early hypothesis. 

\subsection{Exploding probability estimates}

A problem that arose is that in some instances the estimated probabilities $\pi(q)$ became larger than one. This problem arises due to the fact that probabilities that occur in the access sequence of that node get underestimated, forcing $\pi(q)$ to become larger to meet the criterion $P(x_{q}^A) = \pi(x_{q}^A) \cdot \pi(q)$. We want to mention two strategies to deal with this problem. Firstly, the learner can simply continue adding nodes to the observation tree until $\pi(q)\leq 1 \text{ } \forall q \in Q$. We show convergence in section \ref{sec:proofs}.

For the second approach we clip the stopping probability to a maximum value $\pi(q) \leq 1 - \epsilon \text{, } \epsilon > 0$, and continue learning. A drawback of this approach is that we cannot guarantee $P(x_q^A)=\pi(x_q^A)$ to hold anymore.

\subsection{Equivalence testing}

In practice it is not feasible to determine equivalence exactly. To approximate equivalence usually an input string strategy is employed, and the challenge is to have an optimal coverage on tested strings. We resorted to a simple random string testing. Then, if for a current hypothesis $\mathcal{H}$ the equivalence oracle failed to find a counterexample within $10000$ randomly generated strings, we considered the systems equal and output the hypothesis. Albeit simple we yield good results with our method already.

\subsection{Convergence of estimated probabilities}\label{sec:proofs}

\noindent
{
\bf Lemma 1} {\it After each DFS search strategy on the observation tree, the probabilities $\pi(x)$ for each sequence $x$ for which $\tau(x)$ is defined on the observation tree is $\pi(x)=P(x)$.
}

\vspace{\baselineskip}

The proof of this follows by design of the algorithm. The next one gives us some idea about the behavior of the distributions in the limit.

\vspace{\baselineskip}

\noindent
{
\bf Lemma 2} {\it Assuming the teacher returns correct probabilities $P(x)$, $\pi(q_0, a)$ will converge to the real probabilities $P(a\Sigma^*)$ for each $a\in \Sigma$. Furthermore, after initialization of the root node $\pi(q_0)$ remains correct at all times.
}

\vspace{\baselineskip}

The proof follows the simple definition of probabilities. For the root it goes that for a given $a\in \Sigma: \text{ } P(a\Sigma^*)=\sum_{x \in \Sigma^*}P(ax)$. Algorithm \ref{alg:update_path} ensures that this condition is fulfilled as the number of states $n$ in the observation tree grows to $n \rightarrow \infty$. The second statement follows simply from the definition of $\pi(q)$ and that for $q_0$ it is always $x_{q_0}^A=\lambda$. Note that as a side consequence of this we get $\sum_{a \in \Sigma} m(q_0, a) + \pi(q_0)=1$ in the limit.

\vspace{\baselineskip}

\noindent
{
\bf Lemma 3} {\it We again assume the teacher returns correct probabilities $P(x) \text{, } \forall x \in \Sigma^*$. Then, for each node $q \in Q$ of the observation tree with access sequence $x_q^A$ its respective attribute $m(q, a)$ will converge to $P(x_q^A a \Sigma^*)=\sum_{x'\in \Sigma^*}P(x_q^A a x')$, again $ \forall a \in \Sigma$.
}

\vspace{\baselineskip}

Again, the proof follows from the definition. With these two we can now proceed to the following, more interesting theorem.

\vspace{\baselineskip}

\noindent
{
\bf Theorem 1} {\it Given the algorithm as described in this work, the probabilities $\pi(q)$ and $\pi(q, a)$ of each node $q$ of the observation tree converge to the real probabilities as the number of nodes $n$ in the observation tree grows towards the limit, $n \rightarrow \infty$. 
}

\vspace{\baselineskip}

Starting from the root node, we know that the stopping probability $\pi(q_0)$ is correct. We also know that $P(a|q_0)$ is the real probability $P(a\Sigma^*)$ for all $a \in \Sigma$. We choose an arbitrary element $y \in \Sigma$. Then, with $q_0$ modeling the probabilities correctly in the limit we know that for $y$ and the child node $q_1=\tau(q_0, y)$, the stopping probability $\pi(q_1)$ must also be correct, since $P(y|q_0)$ is correct. Now it is to remain to show that the distribution $\pi(q_1, a)$ will converge to its true distribution. We start again with the observation that $\pi(y|q_0) = P(y\Sigma^*)$, i.e. it is the sum of the probability of all strings that start with $y$, and we note that for any $a\in \Sigma$, we have

\begin{equation}\label{eq:conditional_chain}
    P(a|q_1)=P(a|y\Sigma^*)*P(y\Sigma^*).
\end{equation}
 
We further require for $q_1$ that $\sum_{a\in\Sigma}\pi(a|q_1) + \pi(q_1)=1$. We can expand the term with Eq. \ref{eq:conditional_chain} and arrive at

\begin{equation}
    \frac{ \sum_{a\in\Sigma}P(yaX)}{P(yX|q_0)} = 1 - \pi(q_1).
\end{equation}

From this we can see that the normalization operation of Alg. \ref{alg:normalize_node} is correct in the limit. The rest of the proof follows by induction. 

\section{Abbreviations}

\begin{itemize}
    \item SUL: System under learning
    \item P(D)FA: Probabilistic (Deterministic) Finite Automaton
    \item RNN: Recurrent Neural Network
    \item DFS: Depth First Search
\end{itemize}

\section{Notation}

\begin{itemize}
    \item $\Sigma$: Alphabet
    \item $a$: Token in $\Sigma$
    \item $x$: String over $\Sigma^*$
    \item $|x|$: Length of string $x$
    \item $\lambda$: Empty string
    \item $Q$: Set of states of automaton
    \item $q_i$: A state in $Q$, indexed by $i$
    \item $q_0$: A unique initial state in $Q$
    \item $\tau$: Transition function $Q \times \Sigma \rightarrow Q$
    \item $Q_{\tau(q)}$: The set of all reachable states from state $q$
    \item $X_{\tau(q)}$: The set of shortest strings to reach states in $Q_{\tau(q)}$ s.t. each state $q' \in Q_{\tau(q)}$ has exactly one associated $x \in X_{\tau(q)}$: $\tau(q, x)=q'$
    \item $\pi$: Probability $Q \times \Sigma \rightarrow [0, 1]$
    \item $x^A_q$: The unique access sequence of node $q$ in the observation tree
    \item $\mu$: Error bound for learning
    \item $\mathcal{H}$: A hypothesis
    \item $\mathcal{T}$: The target/SUL
\end{itemize}

\vskip 0.2in
\bibliography{sample}

\end{document}